\documentstyle[11pt,sgg,twoside]{article}

\begin{document}

\title{N-body simulations of interactions and mergings in small galaxy 
groups}

\author{E. Athanassoula}
\affil{Observatoire de Marseille, 2 place Le Verrier, 13248 Marseille
cedex 04, France}

\begin{abstract}
In this paper I focus on three topics related to the dynamical evolution of
small galaxy groups, for which the input of N-body simulations 
has been decisive. These are the merging rates in compact groups,
the properties of remnants of multiple mergers, and the evolution of
disc galaxies 
surrounded by one or more satellites. The short dynamical times of compact
groups make it difficult to understand why such groups are observed at
all. N-body simulations have pointed out two possible classes of
solutions to
this problem. The first one proposes that there is on-going
formation of compact groups, 
or that the longevity of the group is due to secondary infall. For the
second class of solutions the longevity of compact groups is due
either to their specific initial conditions, or to a massive common
halo, encompassing the whole group. I discuss here these alternatives,
together with their
respective advantages and disadvantages. I then turn to the structure
of remnants of multiple mergers and 
compare the results of N-body simulations with the
properties of observed elliptical galaxies. Finally I discuss the
dynamical evolution of a disc galaxy surrounded by one or more
spherical satellites.  
\end{abstract}



\section{Introduction}

It is now well established that galaxies are not island universes and
that their dynamical evolution is strongly influenced by their
environment. The complexity of the processes at hand hampers
analytical approaches and makes N-body simulations particularly well
suited for such studies. Thus interactions and mergings of galaxy
pairs have been a favoured goal for simulations (cf. Barnes \&
Hernquist 1992, Barnes 1998 and references therein). It was, however,
necessary to 
wait until computer hardware and software reached an adequate level
before the evolution of small galaxy groups could be simulated  
with sufficient resolution. In the following
I address three problems for which N-body simulations have deepened our
understanding. In section~2 I discuss
the merging rates in compact groups, and present two classes of
possible solutions to the 
problem of short merging times. In section~3 I
present the results of N-body simulations of multiple mergers and
compare the structure of the remnants to that of 
elliptical galaxies. Finally in
section~4 I discuss the dynamical evolution of a
small group composed of a target disc
galaxy surrounded by one or more satellites, as a function of the mass
and orbit of the satellites. I particularly focus on the time
necessary for the companion(s) to spiral to the center of the target,
and that as a function of their mass.

\section{Merging rates in compact groups }
\label{sec:rates}

Compact groups are systems of a few galaxies in a tight
configuration. Shakhba\-zian and collaborators (see Hickson 1997 -
hereafter H97 - and
references therein), Rose (1977), Hickson (1982, 1993) and Prandoni, Iovino
\& MacGillivray (1994) produced catalogs
of such groups. The work of Hickson in particular motivated a number
of observational and theoretical studies, reviewed in H97.

An estimate of their dynamical time $t_d$ can be given as
$t_d=R/V$, where $R$ some characteristic size of the group and $V$ a
characteristic internal velocity. The former is just a few times the
typical size of a galaxy, while the intrinsic three-dimensional
velocity dispersion is of the order of 300 km/sec (Hickson et al. 1992),
giving dynamical times which are only a 
fraction of a Gyr. The survival of compact groups over a Hubble time
thus becomes a problem, which has been addressed by a number of N-body
simulations.  
It was initially proposed that such groups are chance alignments
within loose groups (e.g. Mamon 1986, 1995) or filaments seen edge-on
(Hernquist, Katz \& Weinberg 1995; Ostriker, Lubin \& Hernquist
1995). However the large fraction of member galaxies 
showing morphological signs of interactions (e.g. Mendes de Oliveira
\& Hickson 1994; H97 and references therein; Verdes-Montenegro,
these proceedings) and particularly the large fraction of compact 
groups showing an extended diffuse X-ray emission (H97 and references
therein) argue very 
strongly against this hypothesis. We will thus concentrate here on 
other possibilities and on the input of N-body simulations.

The first simulations (e.g. Carnevali, Cavaliere \& Santagelo
1981, Ishizawa et al. 1983, Ishizawa 1986) showed that
galaxies within compact groups interact and then merge within a very
short time, so that the
final stage of the evolution, which is a single large object, is
reached very fast. As computers
progressed it was possible to use more particles per
galaxy, and also to run more simulations, thus allowing a better coverage
of the corresponding parameter space. Such simulations pointed to two possible
classes of solutions to the merging rate problem in compact groups.

\subsection{On-going formation of new compact groups }

One possible solution  
is that compact groups are continuously forming,
in which case the observed groups would have formed only
recently, the older groups having already merged. This was 
first proposed by Barnes (1989), who suggested that the precursors of
compact groups are loose groups. More recently Diaferio,
Geller \& Ramella (1994) used the results of their N-body
simulations to argue that compact groups can form 
continuously in rich collapsing groups. 
An interesting question - whose answer should shed considerable
light on this problem - 
is how the properties of the rich collapsing groups used by Diaferio,
Geller \& Ramella (1994) affect the properties of the resulting
compact groups, and which subset of the initial conditions of rich
collapsing groups leads to the observed compact groups.

In a somewhat similar vein Governato, Tozii \& Cavaliere (1996) used N-body
simulations to follow the evolution of small galaxy groups, each of
which is initially starting as a spherical over-dense
region. There is initially a collapse, followed by a secondary infall
of the surrounding mass. The latter is substantial in a high density
universe and assures the longevity of the group, contrary to the case
of a low density universe, where the secondary infall is not sufficient. 

\subsection{Extending the merging time }

An alternative solution would be to extend the merging time. One could
start by asking what influences the merging rate and whether some
N-body simulations have
given too short merging times because of the initial conditions they have
used. The first step in that direction was made by Barnes (1985), who
used a variety of initial conditions and showed that a massive common
halo, encompassing the whole group,  generally delays the
mergings. This was later confirmed by further simulations by Bode, Cohn \&
Lugger (1993).

Governato, Bhatia \& Chincarini (1991) explored rather specific initial
conditions and found that the group they simulated lasted 9 Gyrs. Their group
consisted of four galaxies with unequal masses. The two
big galaxies had 75\% of the total mass of the group, the remaining 25\% being
shared equally between the two small galaxies. One can thus describe
the group as
a binary with two small satellites. It is expected that such a group
can last for a very long time for an appropriate orbit of the binary, 
since the satellites
are too small to produce any substantial perturbation, and this was indeed
verified by the simulation of Governato et al. They further tested that
the initial conditions of the satellites did not influence the
results, and that a group of four equal mass galaxies merged
considerably faster, as expected.

Athanassoula, Makino \& Bosma (1997, hereafter AMB97), using a very
large number of 
simulations, set out to determine what influences the merging rate in
compact groups and how. Their simulations started out with five
identical spherical galaxies. The halo was either common to the whole
group (hereafter common halos), or attached to each galaxy individually
(hereafter individual halos). They considered different ratios of
halo-to-total mass, 
different extents of individual halos, different density
distributions in the common halo, different distributions of the
centers of the galaxies and different initial kinematics (i.e. groups
in isotropic virial equilibrium, as well as expanding, collapsing or
rotating groups). In order to be less influenced by the random
distribution of the galaxy centers AMB97 made five realisations of each 
case. This gave them a large number of simulations from which to draw
conclusions, but even so does not cover fully the whole possible
parameter space.

AMB97 found that in general groups with individual halos merge faster
than groups with common halos, and that rotating groups merge slower
than non-rotating ones. They also found that groups with common halos
merge slower if these halos are not too centrally concentrated and if
they have a high halo-to-total mass ratio. In order to see how much
all of this can influence the merging rates, they built a group 
with a high halo-to-total mass ratio and with a common halo which was
not too centrally concentrated, and found that it survives without merging
for much longer than a Hubble time. 

It is thus possible that, for appropriate initial conditions, the
longevity of a compact group is not a problem, thus providing a possible
explanation to why so many compact groups are observed in the local universe. 

\subsection{Advantages and disadvantages of the above solutions }

Let us first consider the scenario in which compact groups are continuously
formed at a sufficient rate to make up for the ones that are rapidly merging.
Its advantage is that a number of observational
studies (H97 and references therein) have shown that compact
groups are often associated with loose groups. Nevertheless this
applies to a fair fraction of the groups, but not all. 
H97 noted two potential problems with this scenario. The first one 
was already raised by Sulentic \& Raba\c{c}a (1994) who, studying the
optical luminosity function of galaxies in compact groups and
comparing it with that of isolated ellipticals, found no population of
field ellipticals that are sufficiently bright to be the product of
the merging of a whole compact group. The second point is that, 
if the merging times are short and new compact groups form
continuously, then there must be a lot of merger remnants around
(Mamon 1986). Would 
that not lead to too many remnants?
Where are such remnants hiding?

Two possible fossils of galaxy groups have been so far
reported. Ponman et al. (1994) observed RX~J1340.6~+~4018 both in
X-rays and optically and find it to have a high X-ray luminosity,
comparable to those of the brighter compact groups, while its optical
properties are typical of those of giant elliptical galaxies. A second
candidate, NGC~1132, was reported by Mulchaey \& Zabludoff (1999),
again with the help of X-ray and optical observations. More
candidates may of course show up. Nevertheless present results indicate that
the number of fossils of compact galaxy groups can not be sufficient
to account for a large number of collapsed groups, if the merging rate is high.

The secondary infall solution depends on the value of $\Omega$, since
it can only work for $\Omega$ of the order of 1. Furthermore in this 
model the merging starts rapidly leading to at least one
merger remnant, but it is effectively terminated when the infall
becomes dominant. Thus it should lead preferentially to compact
groups with one (or a couple of) large elliptical(s) surrounded by
smaller galaxies; 
a picture which is far from being always true in observed Hickson groups.

The solution of specific initial conditions will work only if there is
a binary system dominating the dynamics, which is also not always
the case in observed Hickson groups. 

The solution proposed by AMB97 will work, as
discussed in the previous section,
provided compact groups have common halos which are
sufficiently massive and not too centrally concentrated, and perhaps
appropriate kinematics of the galaxy distribution. 
The existence of heavy common halos
has been well established for many compact groups, either with the help of
X-ray observations (e.g.  
Mulchaey et al 1993, Ponman \& Bertram 1993, Pildis, Bregman \& Evrard
1995) or optical observations (e.g. Perea et al. these proceedings,
H97). They should, in many if not most cases, be
sufficiently massive for the AMB97 solution to hold, 
particularly since the mass of the observed hot gaseous halo
should be added to that of the dark matter, as it shares the
necessary dynamical properties. Thus this solution looks very
promissing. Nevertheless a definite answer can
only be obtained with modelling of individual cases, which must await, on the
observational side, well established radial profiles for the dark
matter, and, on the N-body side, a large number of simulations with
common halos of different masses and radial profiles, and different
kinematics of the galaxy distribution.

First-ranked galaxies observed in compact groups do not appear to be merger
remnants since they are not preferably ellipticals or S0s (H97). This
can be difficult to understand in the framework of the theories of
continuous formation or of secondary infall, but follows easily from
the theories of specific initial conditions and of massive common
halos. Furthermore Zepf \& Whitmore (1991) noted a lack of recent mergers
amongst all compact group galaxies. This is in contradiction with the
theory of continuous formation, but in agreement with the other three.

Finally let me note that it is not necessary that one single solution
explains all  
compact groups. Indeed such groups are known to be a
heterogeneous class of objects, and different solutions might prevail in
different cases.

\section{Multiple merger remnants }
\label{sec:remnants}

In this section I will briefly summarise results 
on the structure of remnants obtained from multiple
mergings. The first to address this problem was Barnes (1989),
followed by Weil \& Hernquist (1994; 1996, hereafter WH96), and Athanassoula 
\& Vozikis (1999 and unpublished, hereafter AV99). I will supplement
the results of these studies with a few more recent and yet
unpublished results of mine.

WH96 and AV99 have complementary approaches. WH96 have few
runs, 6 only for multiple merger cases, but a large 
number of particles, of the order of 800\,000 per simulation. This
allows them to resolve  
structural details in the merger remnants. On
the other hand AV99 have more than 300 runs, but, at least in
some simulations, with few particles. The number of particles is not
the same in all their simulations, 
but depends on the luminous-to-total mass ratio, and ranges between 16\,350 and
250\,000. Thus   
in the simulations with fewer particles they could only calculate bulk 
properties of the merger remnant and not details in its structure. On
the other hand their large number of simulations allowed them to draw
some conclusions about how the global properties of the merger remnant 
depend on whether the halo in the initial compact group was common to
the whole group or distributed around each individual galaxy, on the
luminous-to-total mass in the system, on the central concentration
in the compact group, and on its initial kinematics (virial
equilibrium, expansion, collapse, or rotation). In the study I am
currently pursuing I combine the positive aspect of
the two studies, i.e. the large number of particles of WH96 with the
large number of simulations of AV99.

A further difference between the WH96 and the AV99 studies is that in
WH96 the simulations were preselected so as to ensure a very fast
merging after the beginning of the simulation. This is definitely not the
case for AV99. Thus in WH96 the individual mergings occurred one soon
after another, while in AV99 the time between individual
mergings was variable, in some cases allowing an equilibrium to be
reached before the next merging occurred. This may have an effect
on the properties of the final remnant.

Barnes (1989) showed that the radial profile of the projected density for the 
main body of the remnant could
be well approximated by an $r^{1/4}$ law, in good agreement with
observations of elliptical galaxies and with results obtained by
simulations of merger 
remnants of galaxy pairs (e.g. Barnes \& Hernquist 1992, Barnes 1998
and references therein). This was later 
confirmed by WH96 and AV99. The agreement with observations, however,
does not extend to the 
center-most parts, which, in the simulations, show a core rather than a cusp 
(WH96). The problem is more acute for multiple mergers than 
for pairs (WH96). For the case of pairs the inclusion of bulges in the
progenitor disc galaxies reduces the size of the core region. This is
true also for multiple mergers, but to a lesser extent (WH96). Mihos
\& Hernquist (1994) added a gaseous component to the progenitors in
merging pairs, but
even so could not find radial profiles which are, in their inner parts,
similar to the observed ones, this time because of the formation of an
over-dense inner nucleus. The effect of gas in the progenitors of
multiple mergers on the center-most part of the merger profile has not
yet been explored. 

Another question addressed by these studies is the alignment between
the minor axis and the angular momentum axis. Observations show that
the distribution of the angle between these two axes has a maximum
around 0\deg, with a secondary maximum around 90\deg (Franx, Illingworth \&
de Zeeuw 1991). On the other hand the simulations of merging pairs
(Barnes 1992) show 
misalignment angles inconsistent with this picture. Later
simulations, however, give only small misalignment
angles, arguing that the mass ratio of the progenitors, as well as
their properties, have a significant impact on the
merger remnant (Barnes 1998).
Simulations of multiple mergers (Weil \& Hernquist 1994 and AV99) show 
misalignment angles in good agreement with observations, in all cases where 
rotation was sufficiently large for its angle to be
accurately measured.

\section{Small groups consisting of one disc galaxy surrounded by one
or more satellites} 
\label{sec:satellites}

\subsection{One satellite}

\begin{figure}
\vspace{5.7 cm}
\includegraphics{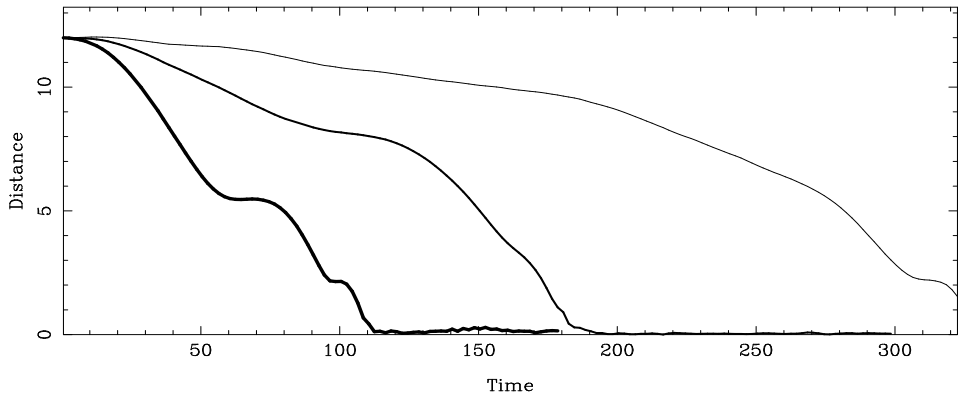}
\caption{Distance from the center of the companion to the center of
the target galaxy as a function of time in systems with only one
satellite. The heavy line corresponds to a
companion of mass equal to that of the disc and the thin one to
a companion ten times less massive. The
line of intermediate width corresponds to a simulation with a
companion of mass 29\% of that of the disc. The unit of length is 3.5
kpc and the 
unit of time is 1.2 $\times 10^7$ yrs.} \label{figure 1}
\end{figure}

I have run a series of simulations with a target disc galaxy and a
companion. The target consists of a disc and a halo with a mass 
ratio of 0.7:1.3, and is bar unstable. I thus first evolved it in
isolation, so that it 
developed a bar, and then added the companion, initially in a
quasi-circular orbit with a radius 
including 96\% of the total mass of the target galaxy. The companion was
in all simulations modeled by a spherical Plummer 
distribution, of the same scale-length and cut-off radius, and its
mass was either equal to that of the target disc,  
or 29\%, or 10\% of it. In this series of simulations the orbital plane 
coincided with the equatorial plane of the target disc and the
number of particles in the target was equal to 800\,000, out of which
280\,000 in the disc. The number of
particles in the companion depends on its mass, so that the mass per
particle is always the same.

In all simulations the companion spirals inwards towards the center of the
target, due to dynamical
friction. The time necessary for 
this depends strongly on both the mass of the companion and
the sense of its rotation around the target. As can be seen in Figure
1, more massive
companions fall in faster, in good agreement with Chandrasekhar's law
of dynamical friction (e.g. Binney \& Tremaine 1987).  
Retrograde orbits lead to much longer 
infall times than direct ones in the case of low mass companions,
while the sense of rotation makes no difference in the case of high
mass companions. 

As the high mass (density) companion spirals to the center of the disc
it looses  
only a small fraction of its mass before it reaches the center of 
the target. There it occupies the place where a bulge would normally
be located. The main change it undergoes concerns its shape, which becomes
oblate, due to the extra gravitational attraction of the target
disc. At the same time the target disc both thickens and expands, so
that its axial ratio is little changed in the process and it still
remains a disc. On the other hand the bar is destroyed by loosing a
lot of its particles and wraps around the companion. Thus the final
radial density profile of the disc
has a minimum in the center, and can be thought of as
a Freeman type II profile.

The fate of a low mass (density) companion is totally different. As 
it spirals inwards it looses a fair fraction of its
mass and its spherical shape, becoming strongly elongated in the
orbital plane. Particles escaping the companion towards the
outer parts form a one-armed spiral 
feature, which with time grows thinner, more tightly wound, expands
and gradually disappears leaving behind it a thick disc. Particles
escaping the companion towards the inner parts develop, in the case of 
a direct companion, orbits elongated along the bar and form mass
concentrations around the two ends of its major axis. In the case of a
retrograde companion 
the particles escaping it towards the inner parts form a ring-like feature
around the bar. The bar itself is not destroyed,
but suffers small changes, both in its pattern speed and amplitude.

The satellite also severely influences the kinematics of the halo,
since it gives it some of its orbital angular momentum. Thus the halo
acquires a sizeable spin in the case of the high mass satellite and a
smaller but still measurable one in the low mass case. As the companion spirals
inwards it also increases locally and temporarily the velocity dispersion 
of the halo. This local maximum spirals inwards to the
center of the halo, together with the companion.

\subsection{Several satellites}

\begin{figure}
\vspace{15.0 cm}
\includegraphics{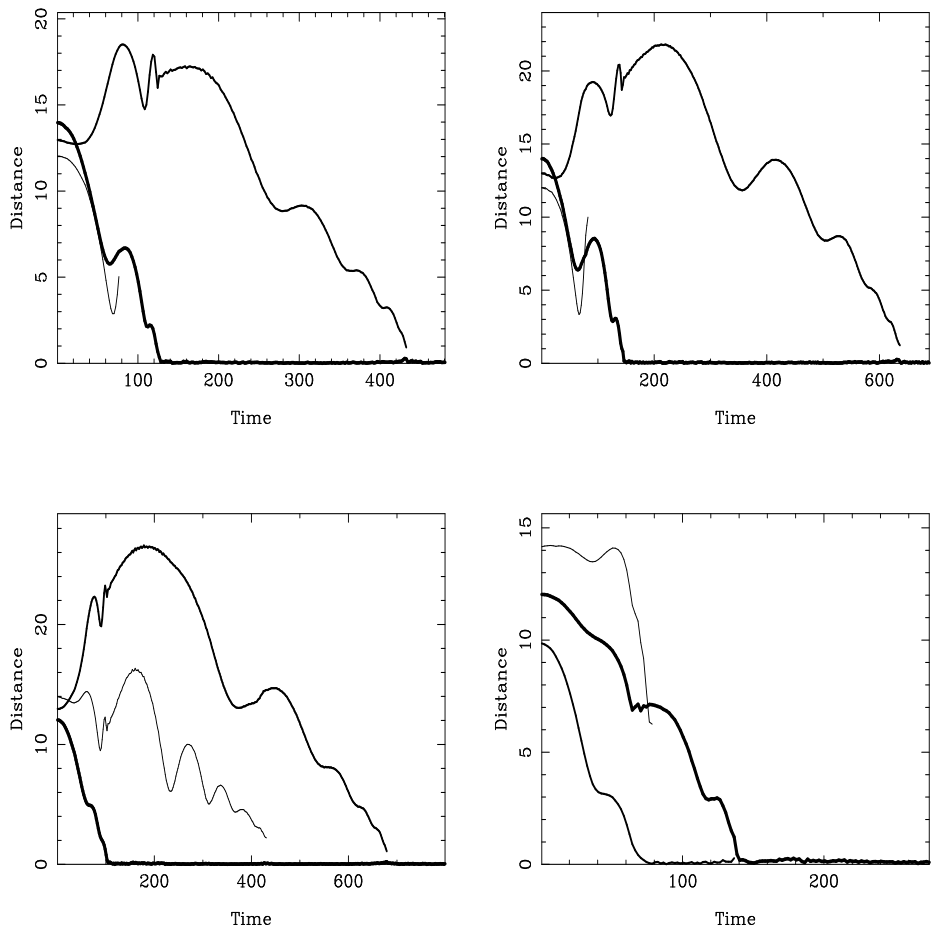}
\caption{Distance between the center of the companion galaxy and the
center of the target as a function of time. Each panel corresponds to
a different simulation, each with three companions of different
masses and initially at different distances from the target. The thickest
line shows the evolution for the companion of mass equal to
that of the disc and the thinnest corresponds to the companion which
is ten times 
less massive. The line of intermediate
thickness is for the companion of mass equal to 29\% of that of the
disc. Units are as for 
Figure 1.} \label{figure 2}
\end{figure}

I have run a few simulations where the target disc galaxy
is initially surrounded by three satellites. In all simulations one
satellite was 
massive (of mass equal to that of the disc), one had a low mass (equal 
to one tenth of that of the disc), and the third one had an
intermediate mass (somewhat less than a third of the mass of the
disc). What changed from one simulation to the other 
was the initial positions of the companions, including their distances
from the center of the target and their relative phases.
Since I am only interested in the time evolution of global quantities,
like the distances between the companions and the target,  
I used relatively few particles per simulation. The 
target was modeled with 120\,000 particles, out of which 42\,000 in
the disc and the remaining
in the halo. The number of particles in the companions was 42\,000, 
12\,000 and 4\,200 respectively.

As was clear from the short summary given in the previous subsection,
simulations with only one satellite give a wealth of interesting
features. It is nevertheless relatively easy to draw general
conclusions and give a global picture of the results. This is much
more difficult to achieve in the case of multi-satellite simulations,
where the evolution depends on many more factors. For example in
the case with only one satellite one could say that the companion
sinks towards the center of the target quasi-monotonically, at a rate
depending mainly on its mass and sense of its orbital rotation.
As can be seen from the four examples
shown in Figure 2, the situation is much more complicated
in the case of three companions, since these interact not
only with the target, but also with each other, exchanging energy and
angular momentum between them. Thus satellites can temporarily move
outwards rather than inwards, or lighter satellites can sink faster
than more dense ones. Furthermore satellites can merge between them
before merging with the target. It is thus not possible to propose a
simple, global picture as in the case of a single companion.

\acknowledgments

I would like to thank Albert Bosma for many useful discussions,
J.C. Lambert for his help with the GRAPE simulations and Philippe
Balard for producing the videos shown during the talk.

\end{document}